# Effects of Syringe Pump Fluctuations On Cell-Free Layer in Hydrodynamic Separation Microfluidic Devices

Md Ehtashamul Haque[1], Amirali Matin[2], Xu Wang[2] and Maïwenn Kersaudy-Kerhoas[1,3*]

[1] Institute of biological Chemistry, Biophysics and Bioengineering, School of Engineering and Physical Sciences, Heriot-Watt University, EH14 4AS, Edinburgh, UK

[2] Institute of Photonics and Quantum Sciences, School of Engineering and Physical Sciences, Heriot-Watt University, EH14 4AS, Edinburgh, UK

[3] Infection Medicine, College of Medicine and Veterinary Medicine, University of Edinburgh, UK

*Corresponding author: Maïwenn Kersaudy-Kerhoas, m.kersaudy-kerhoas@hw.ac.uk

**Abstract**

Syringe pumps are widely used biomedical equipment which offer low-cost solutions to drive and control flow through microfluidic chips. However, they have been shown to transmit mechanical oscillations resulting from their stepper motors, into the flow, perturbing device performance. In this work, unlike previous studies at lower flow rates, we have uncovered that the relative pressure fluctuation plateau from 5mL/h onwards to approximately 2% of the average pressure. Furthermore, we find that absolute pressure fluctuations increase as a non-linear monotonic function of kinematic viscosity at flow rates in the 5-25 mL/h range, while the relative pressure fluctuations peak at 1.25 cSt. Using a novel low-cost coded compressive rotating mirror (CCRM) camera, we investigated the effect of fluctuations in a hydrodynamic microfluidic separation device based on a cell-free layer concept. Using this high-speed imaging set-up, we quantified the cell-free zone width fluctuations at bifurcations. We demonstrated that these fluctuations have the same frequency and amplitude than the syringe pump induced pressure oscillations. Finally, to illustrate that pressure fluctuations degrade the separation efficiency in such devices, we demonstrate using milk samples, instances of particles diverted to undesired outlets. This work provides an insight into the effect of syringe pump fluctuations on microfluidic separation, which will inform the design of microfluidic systems and improve their resilience to pulsating or fluctuating flow conditions.

1. **Introduction**

Microfluidics refers to the manipulation and control of fluids at the micro-scale, typically from a few picolitres (pL) to a few microliters (μL) in miniaturized systems [1]. This field has been developing since the early 1990s and microfluidic technology has been adopted in many life sciences and biomedical experimental and commercial systems, for example, in pH control, rheology, optofluidics, cell analysis, and Point-Of-Care (POC) devices. The steady growth of the microfluidic field can be explained by the benefits of microscale physical effects, which enable precise control of fluids and particles, thanks to stable laminar flow conditions. Thus precise and accurate control of flow rates or pressure, with no disruption, is essential in many microfluidic-based experiments such as droplet generation [2], [3], particle separation [4], [5], cell manipulation [6], [7], sample processing [8], [9], and chemical and biomolecular sensing [10]. Three types of systems are used to control liquid motion in microfluidic devices: (i) hydrostatic pressure and capillary action, which do not require external forces, (ii) pressure control via pressure generators, and (iii) flow rate control via syringe pumps. Syringe pumps are widely used for flow rate control because of their availability, user-friendliness, cost effectiveness, and their capability to control the flow rate across microchannels independently of the fluidic resistance. Additionally, syringe pumps are ubiquitous equipment in medical settings, which makes them particularly suitable for medical applications and medical staff. However, stepper motors inside syringe pumps can be an unwanted source of flow fluctuations in medical applications and microfluidic systems [11]. Flow instabilities can cause an undesirable change of regimes for microfluidic separations [12], [13]. In particular, these fluctuations can be a serious concern for microfluidic applications such as droplet microfluidics where flow fluctuations have been demonstrated to increase polydispersity [11], [14]. As an example, in their droplet microfluidics experiment, Li et al. observed a 13 μm radial variation of 435 μm sized formed jet at the meeting point of the two fluids flowing from different capillaries because of the pulsation of the stepper motor from syringe pump [11]. In their study, they found the calculated frequencies of mechanical oscillation from the syringe pump are consistent with the measured frequency of ripples and they are linearly related with the driving flow rate, similar to the results obtained by Zeng et al. where the authors compared the measured frequencies of the induced pressure fluctuation in their device [15]. The later study further established that the deviation between these two frequencies is less than 10%, and smaller diameter syringes usually provide higher fluctuation however, the amplitude of fluctuation is higher for larger diameter syringes.

Other techniques that are particularly susceptible to pump fluctuations are hydrodynamic separation techniques. Microfluidic-based passive separation techniques use specific channel geometries to separate particles in designated outlets and harvest pure particle-free liquid into remaining outlets. These passive techniques have been applied for cell separation [16]–[18], blood plasma separation [19]–[22], algae separation [23], [24], and other applications [25]–[27]. Flow stability is crucial in these techniques because instabilities can cause particles or cell travelling in the wrong flow streams and thus lead to poor separation efficiency[12], [28].

In passive hydrodynamic separation techniques, microchannels can be designed to take advantage of biophysical effects such as the cell-free layer effect [29], [30], and bifurcation effects [31], [32] as illustrated in Figure 1.A. According to the Fahraeus-Lindqvist law, when a mixture of fluid and particles flows through a tube with a diameter of less than 300 μm, the viscosity starts to decrease with a decreasing tube diameter [32], [33] as a result of particle migration to the centre of the tube [34], [35]. The particle migration can be explained as an effect of two different forces: shear gradient lift force, wall induced lift force. While the shear gradient lift force tries to move the particles towards the walls of the channel, the wall induced lift force acts oppositely and move the particles away from the wall. The net lift force from these two determines the equilibrium position of the particle somewhere in the middle of the channel and therefore a cell-free layer (CFL) forms at the tube wall [36]. The hydrodynamic technique can utilise constriction-expansion regions to create additional lift forces and displace particles to desired streamlines, enabling separation in given outlets. This naturally occurring bifurcation effect is sometimes described as Zweifach-Fung bifurcation law [19], [31], [37]. Utilising these effects ensures the generation of cell-free zones, from where the liquid portion is extracted through daughter channels. The stability of the cell-free zones positioning is crucial to maintain the high separation efficiency of these devices.

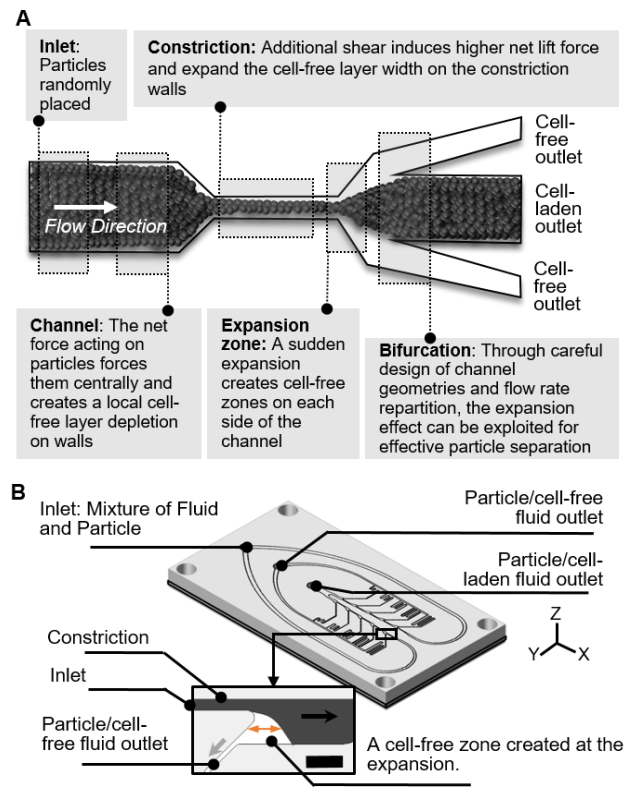

*Figure 1:(A) Graphical illustration of particle, fluid and structure interactions in microfluidic hydrodynamic separation devices (B) The hydrodynamic separation device used in this experiment. Scale bar is 3mm. A similar structure was described in [19]. In inset, the detail of a bifurcation and cell-free zone. Scale bar is 120 μm. The orange arrow indicates where the cell-free zone width (CFZW)*

High-speed imaging has shown an exceptional potential in capturing ultrafast transient phenomena in a variety of chemical and biomedical applications such as high-throughput blood cell screening [38], [39], fluorescence confocal and lifetime microscopy [40], [41] and screening the biological tissues [42], [43] which requires imaging systems with captures rates Kfps to Mfps. In the fields of biomedical research and clinical applications, high-speed imaging facilitates the detection and tracking of cells, and other particles of interest in a specimen individually or as a group with high sensitivity and accuracy. However, bulky design, high cost and complex operation of the conventional high-speed imaging systems make them almost an out-of-reach technology in such resource-limited areas. As a result, accessibility to a portable, low cost and low maintenance high-speed imaging systems is particularly important for such environments.

A novel coded compressive rotating mirror (CCRM) camera can facilitate high-speed imaging of a transient phenomenon at a significantly lower build cost and operation complexity while maintaining a compact and mobile instrument design. Operation principle of the CCRM camera is based on compressive sensing [44] and optical encoding scheme [45] that enables the imaging of non-repeatable events at high speed frame rates in a monochrome and coloured imaging formats.

In this work, using a novel high-speed imaging technique using a coded compressive rotating mirror (CCRM) camera, we set out to investigate the effects of syringe pump fluctuations on hydrodynamic separation structures in particular. To the best of our knowledge, no study has been conducted to observe the effect of flow fluctuations generated by a syringe pump on the cell-free or particle-free layer in hydrodynamic separation devices. The findings from this study were characterized in the context of the separation of milk fat, where the separation efficiency was investigated against the pump fluctuations.

2. **Material and methods**

2.1. Microfluidic chip design and fabrication

The hydrodynamic separation device used in our experiment were previously developed and tested by our group in the context of blood plasma separation [46]. The chip used in this experiment comprises one inlet and two outlets, one outlet for particle-free liquid and one outlet for particle-enriched liquid, as illustrated in Figure 1.B. The main inlet channel separates into two units where each unit comprises five constriction-expansion region, followed by a bifurcation into a daughter channel collecting the particle-free liquid and the particle-laden main channel. The main channel width varies from 100 to 150 μm, whereas the constriction width varies from 35 to 37.5 μm. The chips were fabricated commercially in a proprietary Bisphenol A Novolac epoxy resin (SU8), on a Polymethyl methacrylate (PMMA) substrate (Epigem, UK). A PMMA holder and Polyether ether ketone (PEEK) connectors were machined in-house to enable standard HPLC-type fittings to be connected to luer-lock syringe connections for safe, high-pressure fluid actuation.

2.2. Fluid flow actuation

An overview of the experimental setup is shown in Figure 2A. The sample was injected in the device by a syringe pump (Aladdin Single-Syringe Pump, model: AL-1000HP, pitch S=1.2 mm). With every step generated by the stepper motor inside the syringe pump, the pushing block moves forward, forcing the syringe to flow the liquid in the system. For a specific flow rate, the pushing block covers a fixed distance by a step originated by the motor. The generated pulse during each step has been identified as a source of disturbance in the flow initiated by the syringe pump [11]. Luer-lock plastic syringes (1mL, 4.78 mm ID, Fisherbrand) have been used in all experiments. An inline pressure sensor (LabSmith uPS-series pressure sensor, model: uPS1800-T116-10**) was used to record the pressure fluctuations originating from the syringe pump. All experiments described in this paper were run in triplicate, or otherwise stated.

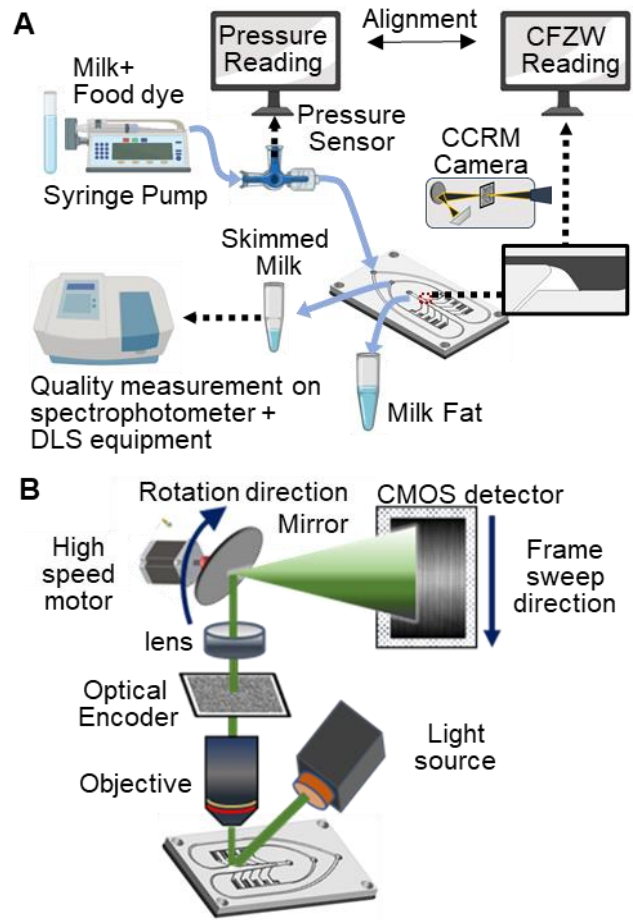

*Figure 2: (A) Microfluidic setup. Samples were injected by a syringe pump. An inline pressure sensor was used to record pressure fluctuations. A high-speed CCRM camera capturing at the rate of 210 frames per second recorded the cell-free zone position for the full run. The pressure sensor data and cell-free zone width data from video file were aligned temporally, and compared to establish the relationship between the syringe pump fluctuation induced pressure fluctuation and cell-free zone width (CFZW). Extract samples collected from the daughter channel were examined with a spectrophotometer and Dynamic Light Scattering (DLS) equipment to study the effect of fluctuation on separation efficiency. (B) Optical (CCRM) set-up. The sample was illuminated using a halogen lamp where the reflected light was collected using a set of objective and relay lens system which was then encoded via an optical encoder. The encoded scene was collected via a lens and focused on the rotating mirror. The reflected light from the mirror was continuously swept across a CMOS detector and was collected by the processing unit for applying the reconstruction algorithm.*

2.3. Optical set-up and image processing

The effects of the fluctuation on the width of the cell-free zone was quantified by using the optical set-up. The optical setup of the CCRM camera [70] is shown in Figure 2.B. The high-speed CCRM camera recorded images of the interrogation window (approximately 400 μm × 350 μm, centred on the first bifurcation) at the rate of 210 frames per second (7× faster that the acquisition rate of the pressure sensor). The dynamic scene was first collected by an infinity-corrected microscopic objective and directed towards the lens tube. The 2D image was focused onto an optical encoder. The encoder pattern was printed on a soda-lime glass with a 1:1 ratio between the blocked and transparent pixels. The encoded image was then collected by an ordinary convex lens and focused on the surface of the rotating mirror. The rotation of the mirror was controlled by an off-the-shelf low cost electric motor [47]. Lastly, the reflected light from the rotating mirror was swept across the surface of the detector (low-cost CMOS module) [47]. The continuous sweeping from the mirror, overlapped the individual encoded frames based on their time of arrival hence applying optical compression on the frames. The optically encoded and compressed observed frame from the detector can be reconstructed via an optimisation algorithm that produces series of individual frames based on the capture rate of the system.

The mathematical representation of data acquisition process using the CCRM camera can be formulated as $y = SPAx + n$, where $y \in R^{MN+(F-1)M \times 1}$ is the captured data by the detector in a vectorized format, $S \in R^{MN+(F-1)M \times MNF}$ is the linear operator of frame shifting and overlapping, $P \in R^{MNF \times MNF}$ is the obtained motion profile of the sweep from the calibration points on the encoder, $A \in R^{MNF \times MNF}$ is the matrix that holds F encoding pattern of $M \times N$ in a diagonal form, $x \in R^{MNF \times 1}$ represents the original frames in a vectorized format, and n is the additive zero mean Gaussian noise. Obtaining x from y is known as an ill-posed linear inverse problem (LIP) [48] i.e., there could be several solutions to the problem hence we transform the acquisition model into a minimization problem and implement the Alternating Direction Method of Multipliers (ADMM) [72] method to solve equations in an iterative fashion. Here, we select the total variation (TV) [73] as the regularizer function where it has demonstrated high performance on various CS based algorithms [74]. One of the main advantages of using TV over the other regularizers is the edge preservation property that prevents hard smoothing of the edge features. This key characteristic prevents the spatial information in in the frames from merging with the background features and therefore averting the loss of the critical information such as the boundaries and intensity amplitudes per pixels that are vital in applications such as high throughput fluidics screening where the boundaries in individual frames are the defining factors in the analysis. Video reconstruction of the original scene can be achieved by solving the formulated minimization problem that is $\hat{x} = arg \min_x \frac{1}{2} ||y - SPAx||_2^2 + \rho_k R(x, \rho_{tv}, w_{tv})$, where $\hat{x}$ is the reconstructed frames that is reshaped into a 3-dimensional data format (video data), $||..||_2^2$ denotes the $l^2$ norm, $\rho_k, \rho_{tv}$ are the variable regularization and denoiser threshold parameters that are adjusted based on the calculated error at each iteration, $R$ is the TV regularization function and $w_{tv}$ is the regularizer weight for the horizontal, vertical and temporal axis.

### 2.4. Sample material

While the chip was designed for blood plasma separation, full fat milk (Tesco Whole Milk (homogenised), 7.4g fat/200 mL) was used in lieu of blood to enable the experiments to be performed at in an optical laboratory with a low biosafety level. Milk and blood share useful properties for the purpose of these experiments. Milk fat takes the shape of large fat globules. Like red blood cells, these globules are soft and deformable globules in a water-based liquid. The composition of milk has been extensively studied. While the average size of fat droplet is 1.6 – 10 μm in non-homogenised milk, following homogenisation, the fat globules size is around 0.2-2 μm, not far from the smallest dimension of red blood cells (RBCs) (2-3 μm in their smallest dimensions). In addition, there are casein micelles (<0.2 μm in size) in the milk as well as various smaller molecules. There are approximately $1.5 \times 10^{10}$ fat globules present in per mL of milk [49]–[51], compared to $4-6.10^9$ red blood cells in plasma. However, milk fat globules and red blood cells significantly differ in density. The density of RBCs is 1.11 g/mL and higher than their surrounding plasma (1.025 g/mL) [52], while milk fat density is 0.9 g/mL and less dense than its surrounding liquid (1.029 g/mL) [53]. Despite their density differences, it is possible to observe milk fat separation in devices designed for blood plasma separation from cells (Figure 1B). Due to the higher shear rate, a particle-free layer is enhanced at the constriction walls. At the expansion-bifurcation, milk fat globules maintain their initial spatial distribution: the fat particles remain on their streamlines and flow into the channel with the higher flow rate. The water-based fat-poor fluid formed at the expansion of the channel, can be harvested through the daughter channel. To observe the behaviour of different dilution samples during experiments, whole milk samples were diluted with Phosphate-buffered saline (PBS, pH= 7 to 7.6, Fisher Scientific) to maintain physiological conditions.

### 2.5. Computational Flow Dynamic (CFD) study

To plot the fluid streamlines, the full device was simulated using the finite element analysis from COMSOL Multiphysics software (v5.3). The 3D separation model designed by AutoCAD (2019 edition) was imported as the device geometry in COMSOL. Water was used as the fluid material but the density and viscosity were changed to 1033 kg/m$^3$ and 2 cP, respectively, to match milk parameters. To save the computational time fine meshing was preferred throughout the whole device which resulted in a total of 525628 domain elements, 142918 boundary elements, and 20355 edge elements. As we are interested in the laminar flow regime, laminar flow physics with two outlets and one pressure driven inlet was adopted and computed with a stationary solver. A no-slip boundary condition was applied to all side walls. After computing, flow streamlines were plotted from specific locations from the internal constriction wall to infer on expected particle locations.

### 2.6. Characterisation of hydrodynamic separation performance via spectrophotometry

The effect of syringe pump pressure fluctuations on the separation efficiency has been characterized using spectrophotometry. A spectrophotometer (Jenway 7315) was used to measure absorbance from 198 to 1000nm wavelengths. The limit of the equipment photometric absorbance is -0.300 to 2.500. To be able to measure in this range, collected milk samples were diluted to 1:250 in PBS before measurement in microcuvette (centre H 15 mm, Brand Gmbh + Co KG) cuvettes. Milk fat was previously reported to absorb strongly at 220 nm [54], [55] in the ultraviolet (UV) range whereas in the Near-infrared (NIR) spectrum milk fat was previously reported to provide a peak at 968 nm [56]. However, absorbance values at 968 nm are very small compared to the 220 nm value because the water content present in milk interferes in this region and strongly absorbs the IR light [57]. Absorbance peaks at both 220 nm and 968 nm were measured from the sample collected from inlet and both outlet sample after 1:250 dilution with PBS solution.

### 2.7. Dynamic Light Scattering

The size distribution of the particles presents in all the collected samples from inlet and both outlets at different flow rates were determined using Dynamic Light Scattering (DLS) (Nanozetasizer, Malvern). All the collected samples were diluted to 1:250 in PBS before the measurements and the analysis were performed at 20 °C.

### 2.8. Statistical analysis

Statistical significance was determined by unpaired parametric Student t-test. Unless specified, p-value significance threshold was set at 0.05. When reporting on statistical significance symbols 'n.s' is used to indicate non-significance (P>0.05), while *, **, *** denotes p<0.005, p<0.001, p<0.0001 as per conventional practice.

## 3. Results

### 3.1. Comparison between of theoretical and measured syringe pump pressure fluctuation frequencies

The goal of this first experiment was to verify the previously established relationship between the measured frequency of the pressure fluctuation and inherent mechanical frequency of the screw-driven syringe pump. The mechanical frequency can be calculated from the following equation, established by [58]:

$$f_m = \frac{4Q}{\pi D^2 S} \qquad (1)$$

Where Q represents the fluid flow rate, D is the internal diameter of the syringe which depends on the model and S is the pitch of the screw in the syringe pump. A 50% milk sample was flown into the device at various flow rates, while the pressure sensor recorded the pressure fluctuations. The frequency of pressure fluctuations was extracted from the experimental data and compared to the theoretical data calculated from Equation (1).

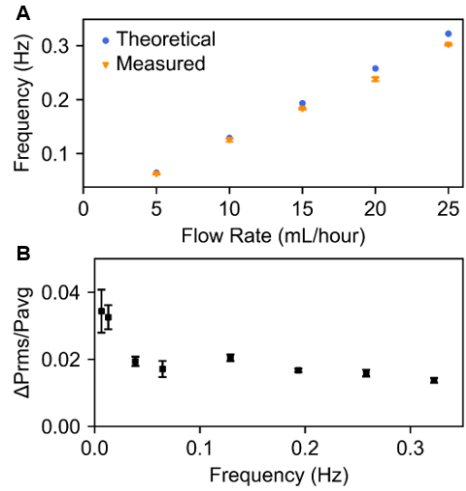

*Figure 3: (A) Comparison between theoretical and measured frequency of syringe pump fluctuations in function of flow rate. The measured frequency was obtained from the pressure sensor measurement while using syringe with an Internal Diameter of 4.78 mm (50% milk sample, N=3). The error on the measured frequency is close to the point size. The theoretical syringe pump frequency is based on Equation (1) for a 1mL syringe, and the pitch of the screw in the syringe pump is 1.2mm. There is no error on the theoretical value of the syringe pump fluctuation frequency (B) Relative fluctuation at various flow rates for 50% milk sample. Up to 0.1 Hz, the linear decay of the relative fluctuation with the frequency corroborates other studies [15]. However, from 0.1Hz (corresponding to 5mL/h) we uncovered that the relative fluctuation reaches a plateau corresponding to a fluctuation with an amplitude around 2% of the average pressure.*

Figure 3.A shows the theoretical and measured frequencies plotted against flow rates ranging from 5 to 25 mL/h. The correlation between the theoretical and measured frequencies is high ($R^2$=0.99). The deviation was more pronounced at higher flow rate. We observed a deviation between theoretical and measured frequencies in the range of 4 to 8%, similar to earlier findings [15]. The relative amplitude of pressure fluctuations was plotted against the pump theoretical frequency in Figure 3.B. Here, we define $P_{avg}$ as the average pressure drop and $\Delta P_{rms}$ as the root mean square (RMS) value of pressure drop fluctuations. Similarly to a previous study, from 0.5mL/h to 5mL/h, the relative fluctuations in our set-up were higher at low flow rates and decreased with an increment in flow rate [15]. However, in previous studies, the highest reported flow rate was 5 mL/h. Higher flow rates are desirable in microfluidic separation devices, as they lead to higher throughput. Thus, we investigated the effects of syringe pump fluctuations at higher inlet flow rates, ranging from 0.5-25 mL/h. Contrary to the previously reported behaviour at lower flow rates, we uncovered that from flow rates >5mL/h (corresponding frequency 0.06 Hz), the relative fluctuation maintains a plateau with an amplitude around 2% of the average pressure. In the case of the 50% milk sample in our device, the pressure fluctuations can reach ±40 KPa (RMS) at 30mL/h. Results for 20 and 10% milk samples are presented in Supplementary Figure S1. Previous studies have shown CFZW at an expansion-bifurcation to increase with the shear rate and the pressure gradient [19], [30]. We hypothesized that the CFZW would fluctuate with the same frequency as pressure gradient fluctuations. In the next section, we investigate experimentally the relationship between syringe pump induced pressure fluctuations and CFZW fluctuations at the expansion-bifurcation.

### 3.2. Relation between mechanical pressure fluctuations and cell-free zone width fluctuations

In the previous section, we have demonstrated that the relative pressure fluctuation reaches a plateau, equal to 2% of the average pressure from 0.06 Hz (5mL/h). To understand the effect of these fluctuations on the separation performance of a hydrodynamic separation device, we must first relate the pressure fluctuations to the CFZW fluctuations, since the CFZW dictates the location of the separation between cell or particle-free liquid and cell or particle-laden liquid, and thus the separation performance. In this section, we investigate the relation between pressure fluctuation and cell-free zone fluctuations in milk samples. Milk samples were flowed through the hydrodynamic separation device at 5, 10, 15, 20 and 25 mL/h and dilution of 10, 20 and 50% with PBS. The pressure and cell-free zone fluctuation were recorded respectively, via the inline pressure sensor and CCRM camera, as presented in Materials and Methods section. Figure 2A shows CCRM-acquired images of the first bifurcation for 15, 20 and 25 mL/h flow rates at 50% milk dilution. The measured average CFZW is indicated for each nominal flow rate.

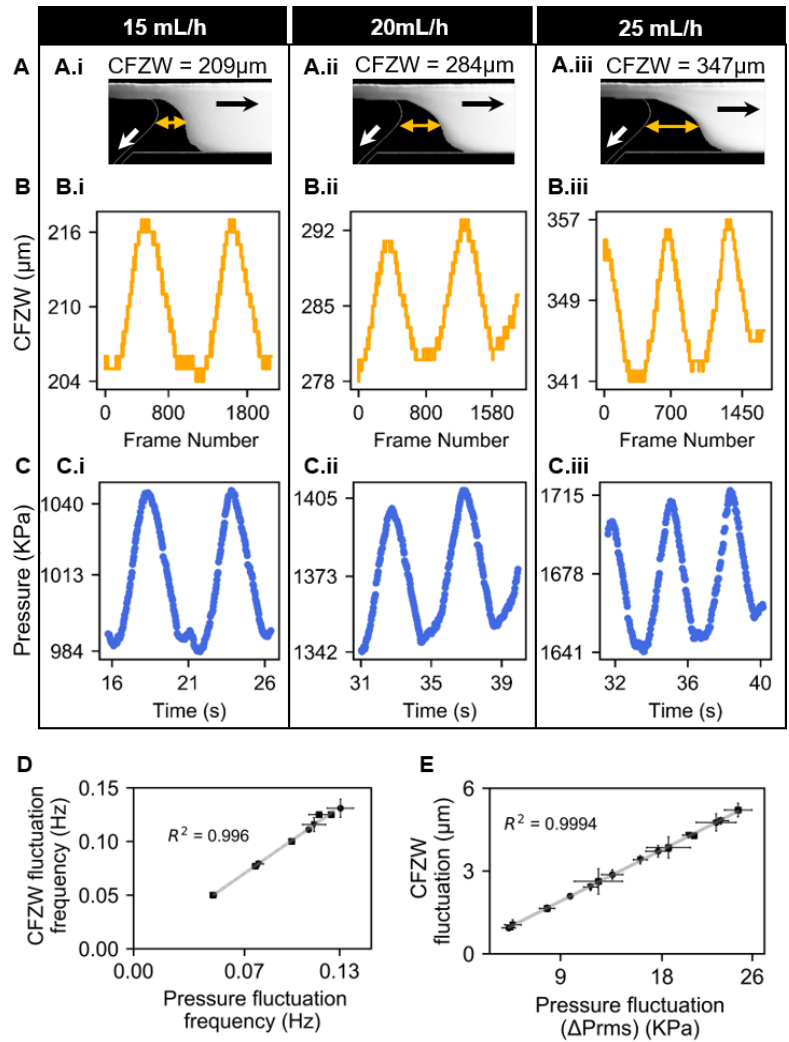

*Figure 4:* **(A)** Photographs of the cell-free zone acquired by the CCRM set-up at 50% milk dilution and 15, 20 and 25mL/h. The yellow arrow indicates the position from which the width of the cell-free zone (CFZW) is acquired. The average of the CCRM-measured CFZW is indicated for each flow rate. **(B)** Raw data for CFZW (CCRM acquisition) and **(C)** Pressure fluctuations (LabSmith pressure sensor, c.f. method section). The pressure sensor provides 30 sample per second where the CCRM camera is able to capture 210 frames per second. The whole data set (including 5, 10, 15, 20 and 25 mL/h flow rate and 1:1, 1:5 and 1:10 dilutions) is available in Supplementary Table S1, S2 and S3 **(D)** Plot of frequency of pressure fluctuations versus frequency of cell-free zone fluctuations for 10% (●), 20% (▲) and 50% (■) milk dilution **(E)** Plot of pressure fluctuations versus cell-free zone width fluctuations for 10% (●), 20% (▲) and 50% (■) milk dilution

The CFZW fluctuations at the first bifurcation are shown in Figure 4.B, while the pressure fluctuations are shown in Figure 4C. The CFZW fluctuations were manually aligned to the pressure fluctuations since the image acquisition and the pressure acquisition could not be automatically synchronised. On these images, we can observe that the CFZW fluctuations reproduce the pressure fluctuations. This can be seen even during minor secondary pressure fluctuations, for example, such as the secondary pressure fluctuations seen between the two peaks in Panel B and C for 15mL recordings, which could be due to temporary minor blockages in the device.

It has been previously shown that PDMS and other rubber-based material of low Young Modulus are known to have damping properties, thereby reducing the amplitude of mechanical oscillations in a fluidic network [60], [61]. It should be noted that here both the chip and the tubing material are non-silicone; the chip is a composite of SU8 and PMMA and the tubing is polytetrafluoroethylene (PTFE). Therefore, most of the hydrodynamic instabilities are expected to be transmitted in the fluid and visible at the CFZW.

As the pressure fluctuations increased in frequency and amplitude at higher flow rates, we observed that CFZW fluctuations also increased in frequency and amplitude. To investigate the correlation between the two sets of data, we first plotted the measured frequencies of the pressure fluctuations against the measured frequencies of the cell-free zone fluctuations. Figure 3.D shows a large correlation ($R^2=0.994$) and agreement between the frequencies of the CFZW and pressure fluctuations. Furthermore, we investigated if the pressure and cell-free zone fluctuations also correlated in amplitude. Figure 3.E illustrates that cell-free zone amplitude is linearly related with the pressure fluctuation amplitude ($R^2=0.977$). Thus, it becomes possible to back-calculate a particular CFZW fluctuation for a given pressure fluctuation amplitude. For example, for a 10 KPa pressure fluctuation from the mean pressure, the CFZW will vary about 2 μm (this holds for 10-50% milk dilution). Since the best in-silico blood models, including the most recent Lattice-Boltzmann models, cannot yet resolve high haematocrit blood flow (or high-density particulates) or support the design of complex, high density soft particulate flow, these experimental findings could be useful when designing hydrodynamic separation devices. By knowing the cell-free zone spatial fluctuations, designers and experimentalists can adjust devices to avoid cells or particles non-intentionally reaching undesired outlets.

### 3.3. Effect of overall sample kinematic viscosity on pressure fluctuations

In the previous section we have related the pressure fluctuation amplitude to the amplitude of the CFZW fluctuations. The pressure gradient inside a microchannel is directly proportional to the viscosity which varies from one sample fluid to another [62]. Therefore, samples having higher viscosity are expected to require higher pressure gradient to flow through the microfluidic network which will result in higher cell-free zone fluctuation which will could further degrade the separation efficiency. In this regard we investigated the effect of pressure fluctuations on samples having different density and viscosity. Since it was impossible to control both viscosity and density independently, these results are reported for kinematic viscosity (ratio of dynamic viscosity over density). The calculated values of densities, dynamic and kinematic viscosities, and are found in Supplementary Table 4.

In a first instance, pressure fluctuations recordings were measured for three milk dilution sample (10, 20 and 50%) in the 5-25 mL/h range with a 5 mL/h increment. We observed the absolute pressure and cell-free zone fluctuations for 50% milk was occasionally marginally superior to that of 20% and 10% milk (Supplementary Figure S2.A and B). As the kinematic viscosity of these three samples were very close to each other, it was difficult to determine clear trends. Therefore, the same experiment was performed with an additional four samples, three diluted glycerol samples (6.7, 20, 33%) and water at 10 mL/h. The variations of pressure were found to increase monotonically with kinematic viscosity as expected (Supplementary Figure S2.C), however a fully linear relationship was not evident ($R^2=0.74$). A similarly poor linearity was observed at 5mL/h (data not shown). Simulation using the compressible fluid module were carried out using the calculated viscosity and density. The deviations between the simulated results and the measured results become apparent at the lower and higher end of the kinematic viscosity spectrum (water and glycerol samples) (Supplementary Figure S3.A and S3.B ). When taken independently, the milk samples provided a good linearity and high correlation and good agreement between the simulated and measured results ($R^2=0.94$) (Supplementary Figure S3.C). The source of the deviations in glycerol samples is unclear. Given the aforementioned non-linearity in the fluctuation, the relative fluctuation amplitude for the same points was found to increase sharply between the kinematic viscosity of 1 cSt to 1.25 cSt, reaching 2% of average pressure. However, after 1.25 cSt, the relative fluctuations started to decrease and maintained a plateau at 1.5% of the average pressure (Supplementary Figure S.2.D).

3.4. Effect of syringe-pump pressure fluctuations on hydrodynamic separation performance

We have established that the pressure fluctuations of the syringe pump introduce fluctuations in the CFZW, and both are linearly related in frequency and amplitude. In a hydrodynamic device, the fluctuations of the CFZW at the bifurcation might become large enough to drive particles in the cell-free outlet and hence reduce the separation efficiency of the device. Two different scenarios are possible: (i) The CFZW is very large and situated away from the daughter channel collecting particle-poor or particle-free fluid. In this case fluctuations in the CFZW are unlikely to lead to any particles traveling to the skimmed milk outlet. The separation efficiency will remain unaffected by the pressure fluctuations (ii) The CFZW is relatively small and situated closely to the daughter channel. In this case pressure fluctuations are most likely to affect the particles path, deviating them from their original position leading to the particle-rich channel into an undesired position, leading to the particle-poor channel. We illustrate both of these theoretical situations in Figure 5.A.

To analyse these scenarios, three distinct flow rates were identified: a lower flow rate (10 mL/h), to investigate the first scenario and two higher ones (20 and 30 mL/h) to investigate the second scenario. In addition, CFD simulation was run for the three flow rates. In this study, the primary focus was of the homogenised fat globules (typically 0.2-2 µm in size) present in the homogenised milk, although some unhomogenised and larger fat globules (~2-6 µm) are also expected to be present in the sample. In a rectangular channel, depending on the aspect ratio (Channel width/Channel depth) and Reynolds number, particle distribution pattern might lead to two to eight equilibrium positions near the walls [64]–[67]. As the Reynolds number in our case varies within the range of 100-300 and the aspect ratio is nearly two, for simplicity we assumed there will be four equilibrium position near the top-bottom and side walls in our case similar to [68]. Therefore, when predicting the positioning of particles at different distances from the side wall, we considered those particles positioning in the z direction (Channel depth) as well. Regardless of the presence of a cell-free layer, particles larger than 2 µm, cannot have their centre of mass between the streamline located at 1 µm from the wall. Smaller fat particles (0.2-0.5 µm) and casein particles (~100 nm) located within the 1.5 µm zone are expected to travel to the skimmed milk outlet, as the net lift force will be negligible for them. Therefore, flow streamlines were computed from precise starting locations from the constriction boundary side wall (1.5, 2.5 and 3.5 µm) at the closest to the daughter channel and for two z location: one at channel centre and another one 4 µm away from the centre.

For the mid-plane scenario ($z=10$), the CFD simulations in Figure 5.B. showed that for any flow rates (10, 20, 30 mL/h), all particles situated 1.5 µm from the edge of the constriction will travel into the particle-free daughter channel (red traces). Whereas, particles situated from and beyond 2.5 µm (black trace) will travel to the particle-rich outlet except the minimum pressure scenario for 10 mL/h when a "switching" event (indicated by an arrow on Figure 5.B) was observed. At this minimum pressure fluctuation for 10 mL/h, the streamlines starting 2.5 um from the wall, switch from the particle–rich channel to the particle-poor channel. A similar switching event at 10 mL/h (minimum pressure fluctuation) was observed when the streamlines were computed 4 µm away from the z-mid-plane and 3.5 µm away from the constriction side wall. For this z plane, all the particles reside within this 3.5 µm distance follows the particle-poor channel, no matter what the flow rate is. This central shift of particles flowing through the particle rich channel is expected as the particles tend to move towards the channel centre near the top and bottom walls to attain their equilibrium position. It can be inferred from the CFD simulation that, although some fat particles will end up at the skimmed milk outlet because of insufficient cell-free zone, the effect of pressure fluctuation which will further degrade the separation efficiency.

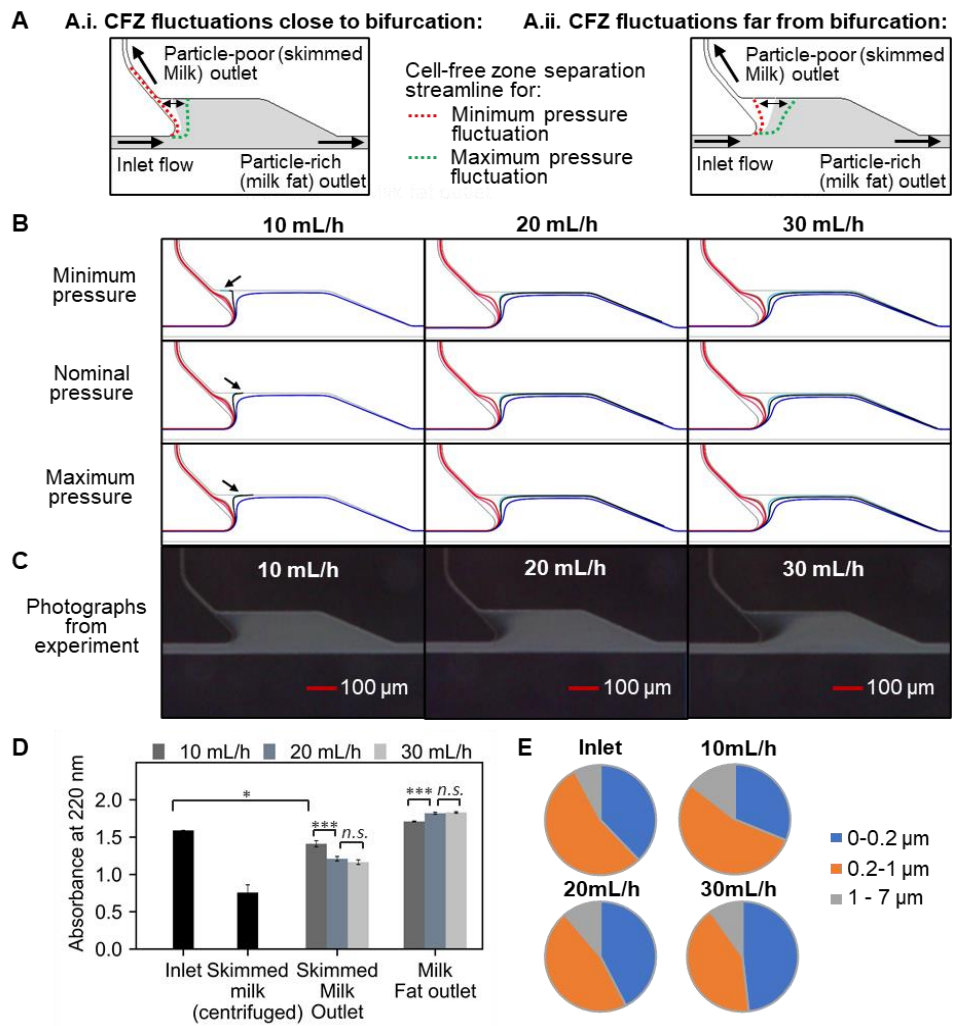

*Figure 5: (A) Schematic explaining the effect of cell-free zone variation on separation efficiency (i) When the cell-free zone resides very close to the skimmed milk outlet at lower flow rates, fat particles have chances to travel to the skimmed milk outlet at minimum pressure fluctuation amplitude and degrade the separation efficiency (ii) When the cell-free zone resides far away from the skimmed milk outlet, fat particles mainly travel to the milk fat outlet even though the fluctuation is higher at higher flow rates (B) computed streamline which are: --- 1.5 µm , --- 2.5 µm and --- 3.5 µm far from the constriction upper boundary are presented for 10,20 and 30 mL/h. The dark coloured streamlines are simulated from the channel centre (Z=10 µm, 20 µm total depth) and light coloured ones are 4 µm bottom from the channel centre (Z=6 µm) At 10 mL/h, particles which are present at the midplane (Z=10) and 2.5 µm far away from the constriction upper boundary have tendency to travel to the skimmed milk outlet when the pressure fluctuation amplitude is minimum. For a different depth (Z=6), this distance can go up to 3.5 µm. Whereas, for 20 and 30 mL/h, fat particles reside at least around 2 µm (For Z=10) and 3 µm (For Z=6) far away from the upper boundary have no chance to travel to the skimmed milk outlet. (C) Photographs of the cell-free zone at 10,20 and 30 mL/h. (D) Absorbance values for the sample collected from Inlet, skimmed milk (centrifuged at 16,000g for 10 minutes), skimmed milk outlet and milk fat outlet at 220 nm (E) Dynamic Light Scattering (DLS) data from the skimmed milk outlets show that the milk fat fraction (1-7 µm)*

*is increased in the 10m/h experiments. However, due to the DLS data variability, statistical significance was not fully achieved.*

Next, to further verify the effect of pressure fluctuations on the separation performance, milk (50%) was flowed into the device at 10, 20 and 30mL/h. Absorbance measurements were acquired at 220 and 968 nm to characterise the presence of fat particles in particle-poor (skimmed milk) microfluidic extract and particle-rich (fat) microfluidic extract. All the absorbance measurements were acquired in triplicates on samples collected from extractions in three unique and previously unused, but identical, devices. For these experiments, the cell-free zone was imaged under a microscope (Dinolite, Premier AM7013MZT) (Figure 5.C) and the absorbance spectrophotometry was used to characterise the separation performance on the inlet (feed) solution. The absorbance values at 220 nm are shown in Figure 5.D for the inlet, and particle-poor (skimmed milk) and particle-rich (fat) outlets at various flow rates. The absorbance of separated centrifuged skimmed milk (at 16,000×g for 10 minutes) is also added as a control. In these experiments, the particle-poor, skimmed milk outlet sample absorbance for 10 mL/h was superior compared to the ones from 20 and 30 mL/h (respectively p= 0.0056, and p=0.024). Although the CFZW was large enough to be situated beyond the particle-poor channel at the nominal pressure, the pressure fluctuations led to more milk fat particles traveling into skimmed milk outlet at 20 and 30 mL/h, the absorbance values remain identical (p=0.17). The values for 968 nm and the full spectrum (190-1000 nm) for all three flow rates can be found in Supplementary Figure S4.A and S4.B. The 968 nm values (Figure S4.C) corroborate the findings at 220 nm. Finally, the inlet and the extracted particle-poor samples were analysed by Dynamic Light Scattering. The DLS results (Figure 5.E.) showed a relatively larger portions of the larger milk fat globules in the 10 mL/h separation, which confirmed the spectrophotometric measurements.

4. **Conclusions**

The effect of syringe pump fluctuations on the cell-free zones in microfluidic hydrodynamic separation structure has been investigated. Using a CCRM imaging approach, we have reported for the first time, a direct relationship between the mechanical oscillation of the syringe pump and the width of the cell-free zone at an expansion region. We have observed that the mechanical fluctuations of the syringe pump cause the cell-free zone width at an expansion region to fluctuate with the same frequency, and amplitude. To explore this effect in the context of high throughput applications, relevant to cell separation studies, the relative fluctuations of pressure and cell-free zone width have been measured in this study from moderate to high flow rates (from 5 to 25 mL/h). We have established that the relative pressure and cell-free zone fluctuation maintains a plateau with an amplitude around 2% of the average value for this range. Finally, we have demonstrated the potentially damaging impact of pressure fluctuations on hydrodynamic separation using milk fat separation. We proved by simple simulation and experimentally, that these pressure fluctuation degrades the particle flow path, and the separation efficiency.

We have also shown that the absolute pressure fluctuations increase with the kinematic viscosity which indicates the pressure fluctuations effects will become more severe for blood or other two-phase, or highly viscous flows.

In conclusion, following these results, microfluidic designers may take two approaches. On one hand, using pulseless syringe pumps or damping solutions will overcome major fluctuation effects and improve the efficiency of separation devices. On the other hand, it is possible to take pressure fluctuations into account and design hydrodynamic separation system for widely used conventional syringe pumps. Using the present findings, designers of bulk cell separation devices can ensure robust performances, independent of pressure fluctuations. Finally, this study could open-up a practical approach for experimentalists wanting to simulating oscillating flows, for example in the context of rheological or haematology studies.

5. **Author Contributions**

MEH, AM, XW, MKK designed the study. MEH and AM performed the experiments. MEH, AM, XW, MKK contributed to the data analysis. MEH, AM, XW, MKK wrote the paper. MKK and XW acquired funding.

6. **Funding**

UK Engineering and Physical Sciences Research Council MicroTotal Pre Analytical Systems (MTPAS): Near-patient Approach to the Preparation of Circulating Biomarkers for Next-Generation Sensing EP/R00398X/1. The imaging research is partially supported by the Scottish Universities Physics Alliance (SUPA).

7. **Acknowledgements**

MEH have been funded by a James Watt Scholarship. MKK acknowledges funding from the UK Engineering and Physical Sciences Research Council MicroTotal Pre Analytical Systems (MTPAS): Near-patient Approach to the Preparation of Circulating Biomarkers for Next-Generation Sensing EP/R00398X/1. XW and AM acknowledge SUPA for support. We would like to thank Noman Naeem, for help with spectrophotometry and Graeme Whyte for technical discussions.

# Effects of Syringe Pump Fluctuations On Cell-Free Layer in Hydrodynamic Separation Microfluidic Devices

Md Ehtashamul Haque, Amirali Matin, Xu Wang and Maïwenn Kersaudy-Kerhoas

Figures

# Figure 1

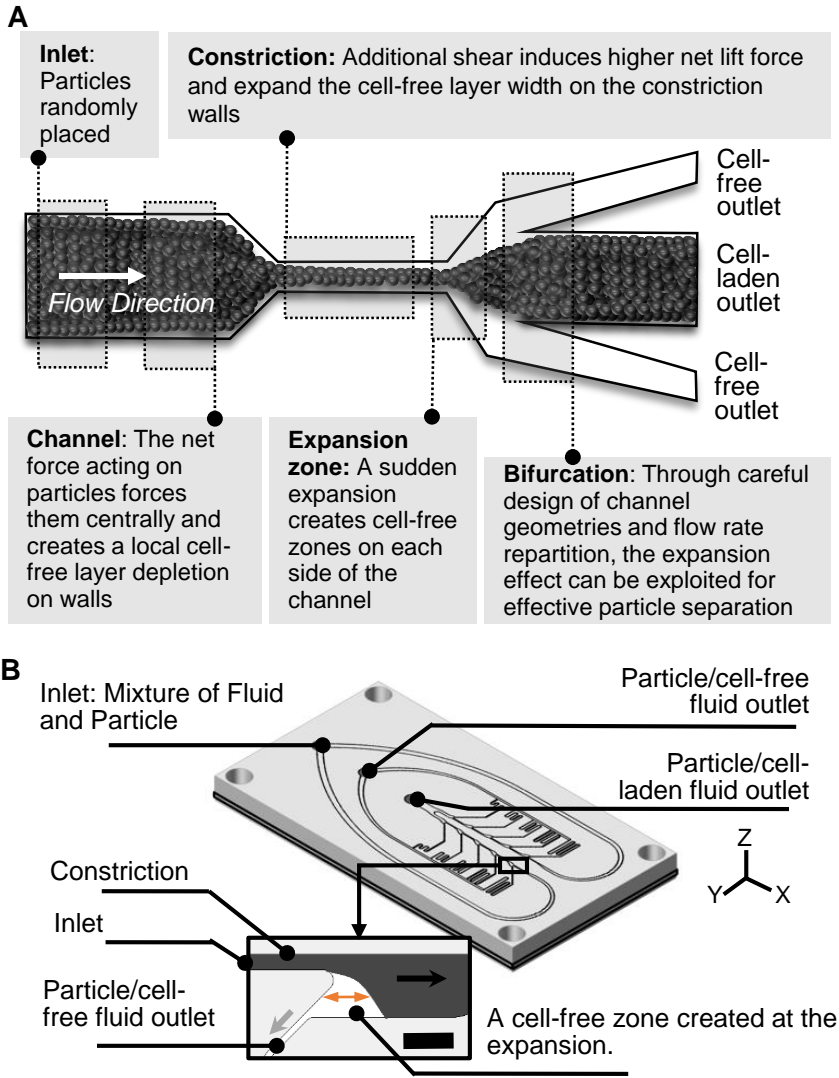

*Figure 1:(A) Graphical illustration of particle, fluid and structure interactions in microfluidic hydrodynamic separation devices (B) The hydrodynamic separation device used in this experiment. Scale bar is 3mm. A similar structure was described in [19]. In inset, the detail of a bifurcation and cell-free zone. Scale bar is 120 μm. The orange arrow indicates where the cell-free zone width (CFZW)*

# Figure 2 (single column)

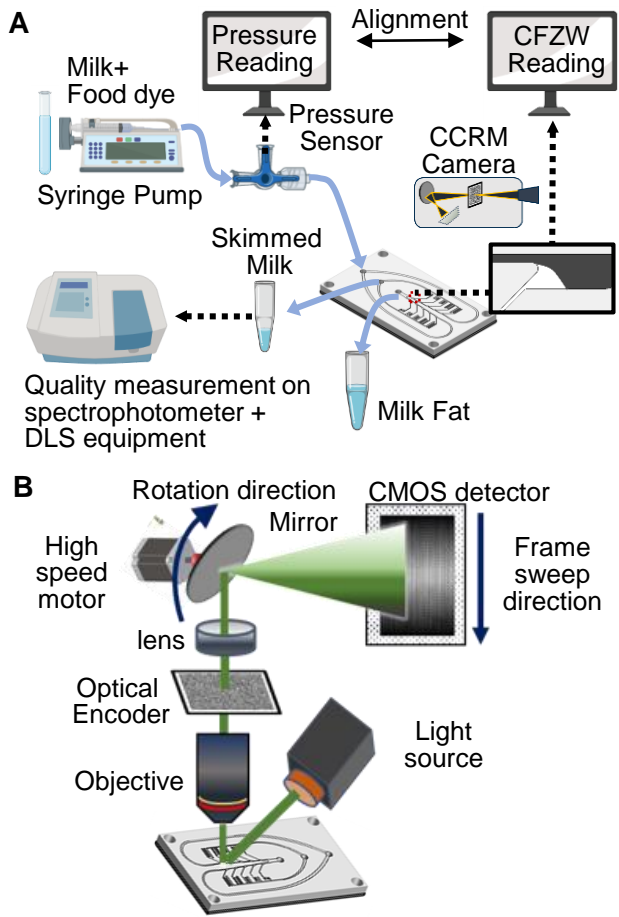

*Figure 2: Experimental Setup: **(A) Microfluidic setup.** Samples were injected by a syringe pump. An inline pressure sensor was used to record pressure fluctuations. A high-speed CCRM camera capturing at the rate of 220 frames per second recorded the cell-free zone position for the full run. The pressure sensor data and cell-free zone width data from video file were aligned temporally, and compared to establish the relationship between the syringe pump fluctuation induced pressure fluctuation and CFZW. Extract samples collected from the daughter channel were examined with a spectrophotometer and Dynamic Light Scattering (DLS) equipment to study the effect of fluctuation on separation efficiency. **(B) Optical (CCRM) set-up.** The sample was illuminated using a halogen lamp where the reflected light was collected using a set of objective and relay lens system which was then encoded via an optical encoder. The encoded scene was collected via a lens and focused on the rotating mirror. The reflected light from the mirror was continuously swept across a CMOS detector and was collected by the processing unit for applying the reconstruction algorithm.*

.

Figure 3

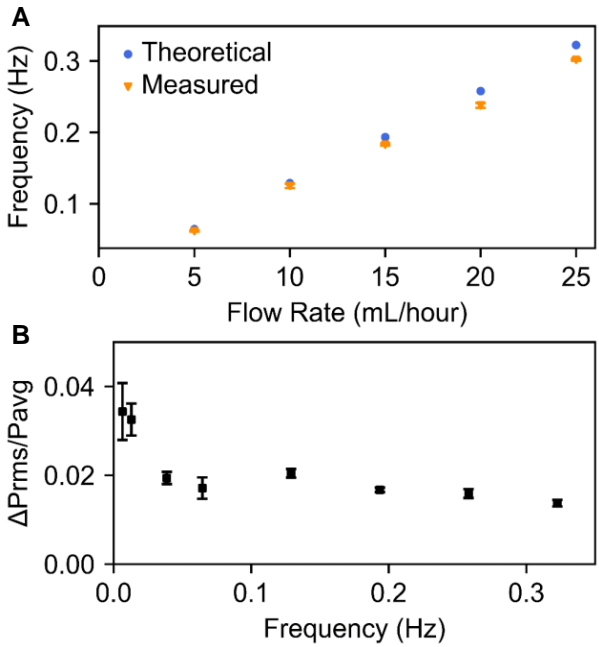

*Figure 3: (A)* Comparison between theoretical and measured frequency of syringe pump fluctuations in function of flow rate. The measured frequency was obtained from the pressure sensor measurement while using syringe with an Internal Diameter of 4.78 mm (50% milk sample, N=3). The error on the measured frequency is close to the point size. The theoretical syringe pump frequency is based on Equation (1) for a 1mL syringe, and the pitch of the screw in the syringe pump is 1.2mm. There is no error on the theoretical value of the syringe pump fluctuation frequency *(B)* Relative fluctuation at various flow rates for 50% milk sample. Up to 0.1 Hz, the linear decay of the relative fluctuation with the frequency corroborates other studies [15]. However, from 0.1Hz (corresponding to 5mL/h) we uncovered that the relative fluctuation reaches a plateau corresponding to a fluctuation with an amplitude around 2% of the average pressure.

Figure 4

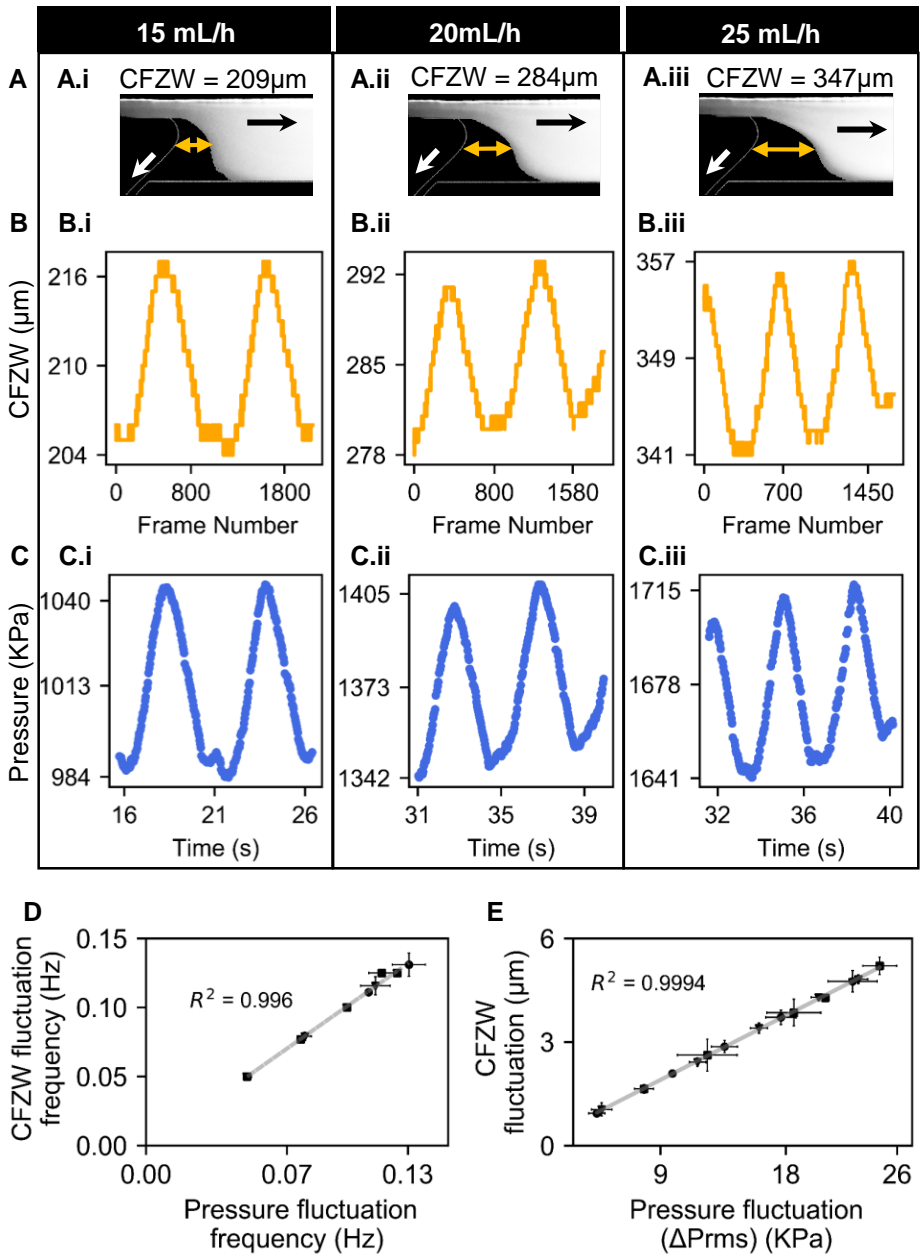

*Figure 4: (A) Photographs of the cell-free zone acquired by the CCRM set-up at 50% milk dilution and 15, 20 and 25mL/h. The yellow arrow indicates the position from which the width of the cell-free zone (CFZW) is acquired. The average of the CCRM-measured CFZWis indicated for each flow rate. (B) Raw data for CFZW (CCRM acquisition) and (C) Pressure fluctuations (LabSmith pressure sensor, c.f. method section). The pressure sensor provides 30 sample per second where the CCRM camera is able to capture 210 frames per second. The whole data set (including 5, 10, 15, 20 and 25 mL/h flow rate and 1:1, 1:5 and 1:10 dilutions) is available in Supplementary Table 1, 2 and 3 (D) Plot of frequency of pressure fluctuations versus frequency of cell-free zone fluctuations for 10% (●), 20% (▼) and 50% (■) milk dilution (E) Plot of pressure fluctuations versus cell-free zone width fluctuations for 10% (●), 20% (▼) and 50% (■) milk dilution*

# Figure 5

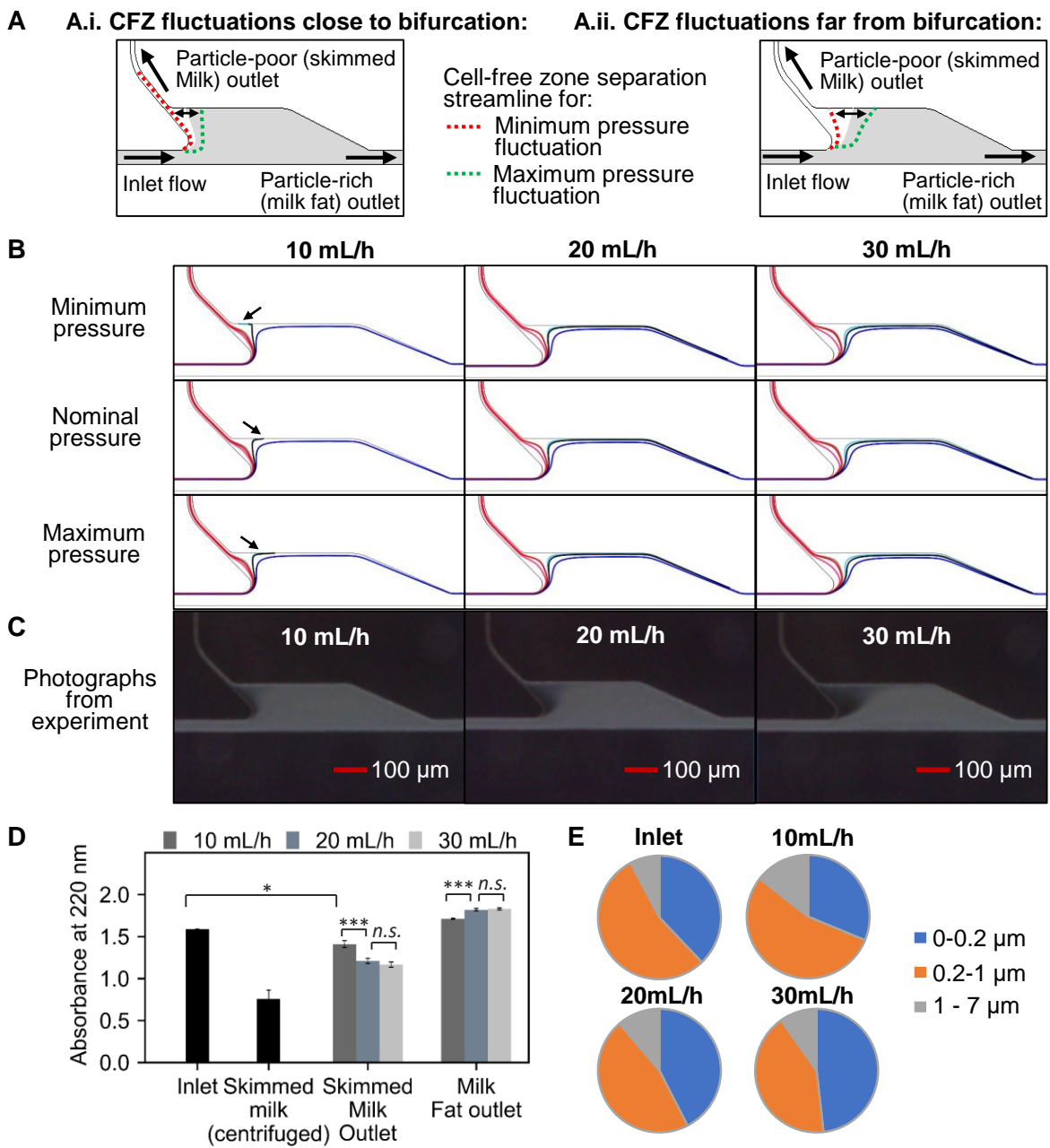

*Figure 5: (A)* Schematic explaining the effect of cell-free zone variation on separation efficiency (i) When the cell-free zone resides very close to the skimmed milk outlet at lower flow rates, fat particles have chances to travel to the skimmed milk outlet at minimum pressure fluctuation amplitude and degrade the separation efficiency (ii) When the cell-free zone resides far away from the skimmed milk outlet, fat particles mainly travel to the milk fat outlet even though the fluctuation is higher at higher flow rates *(B)* computed streamline which are: --- 1.5 µm , --- 2.5 µm and --- 3.5 µm  far from the constriction upper boundary are presented for 10,20 and 30 mL/h. The dark coloured streamlines are simulated from the channel centre (Z=10 µm, 20 µm total depth) and light coloured ones are 4 µm bottom from the channel centre (Z=6 µm)   At 10 mL/h, particles which are present at the midplane (Z=10) and 2.5 µm far away from the constriction upper boundary have tendency to travel to the skimmed milk outlet when the pressure fluctuation amplitude is minimum. For a different depth (Z=6), this distance can go up to 3.5 µm. Whereas, for 20 and 30 mL/h, fat particles reside at least around 2 µm (For Z=10) and 3 µm (For Z=6) far away from the upper boundary have no chance to travel to the skimmed milk outlet. *(C)* Photographs of the cell-free zone at 10, 20 and 30 mL/h. *(D)* Absorbance values for the sample collected from Inlet, skimmed milk (centrifuged at 16,000g for 10 minutes), skimmed milk outlet and milk fat outlet at 220 nm *(E)* Dynamic Light Scattering (DLS) data from the skimmed milk outlets show that the milk fat fraction (1-7 µm) is increased in the 10m/h experiments. However, due to the DLS data variability, statistical significance was not fully achieved.

# Effects of Syringe Pump Fluctuations On Cell-Free Layer in Hydrodynamic Separation Microfluidic Devices

Md Ehtashamul Haque, Amirali Matin, Xu Wang and Maïwenn Kersaudy-Kerhoas

Supplementary Figures

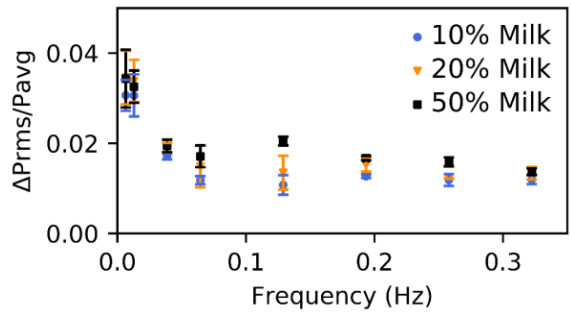

***Supplementary Figure S1:*** *Relative fluctuation at various flow rates for10, 20 and 50% milk sample. In accordance to previous study, the relative fluctuation is higher at lower flow rate* [17]. *However, from 0.1Hz (corresponding to 5mL/h) we uncovered that the relative fluctuation reaches a plateau corresponding to fluctuation with an amplitude around 2% of the average pressure.*

***Supplementary Table 1:*** *Raw data for CFL width (CFZW) (CCRM acquisition) and pressure (LabSmith pressure sensor acquisition) for 50% milk dilution. The dataset were aligned manually, since the image acquisition and the pressure acquisition could not be automatically synchronised. However, the CFL width perfectly mirror the pressure sensor acquisition. The pressure sensor provides 30 sample per second where the CCRM camera is able to capture 210 frames per second.*

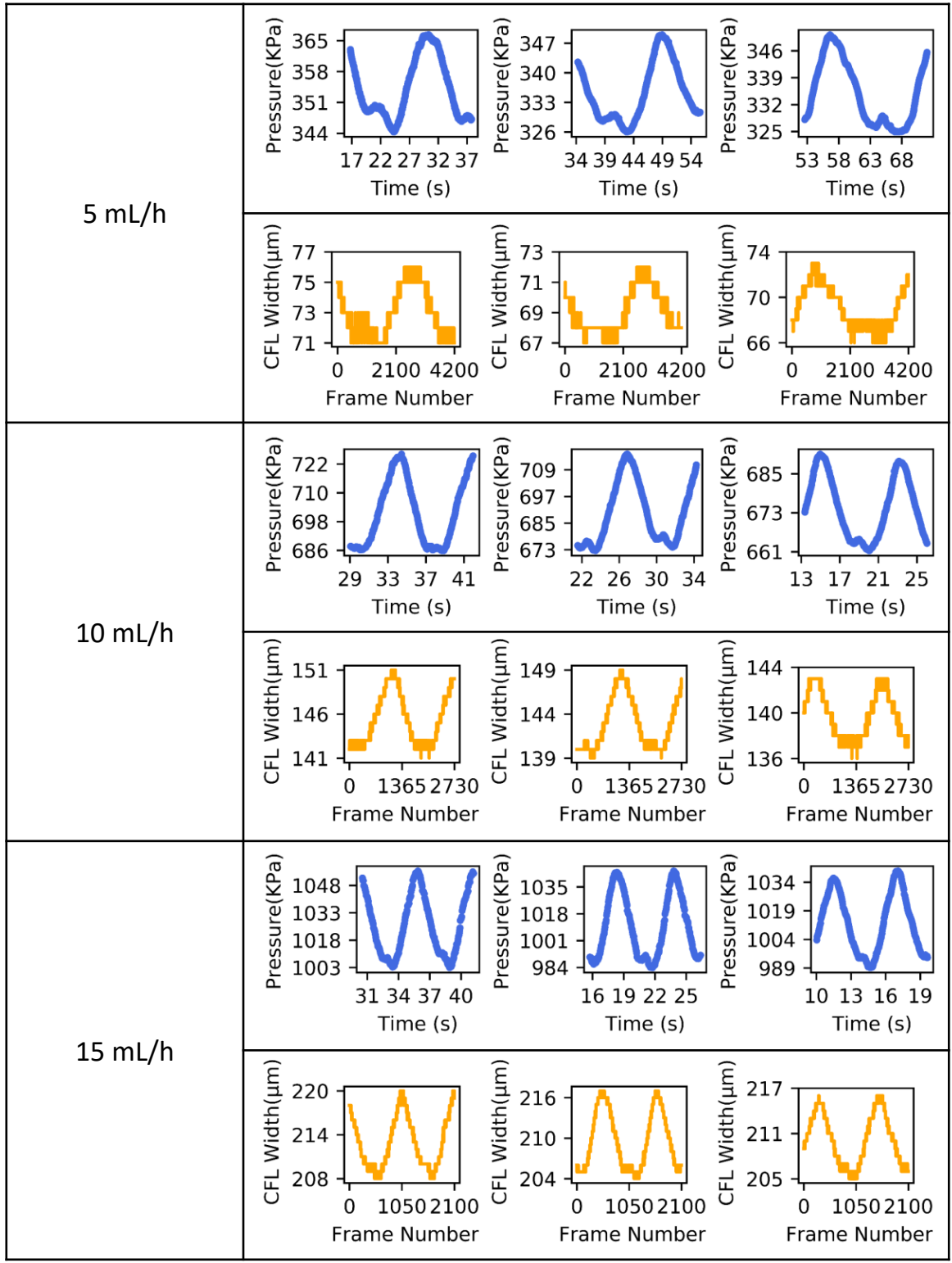

| | |
|---|---|
| 20 mL/h | 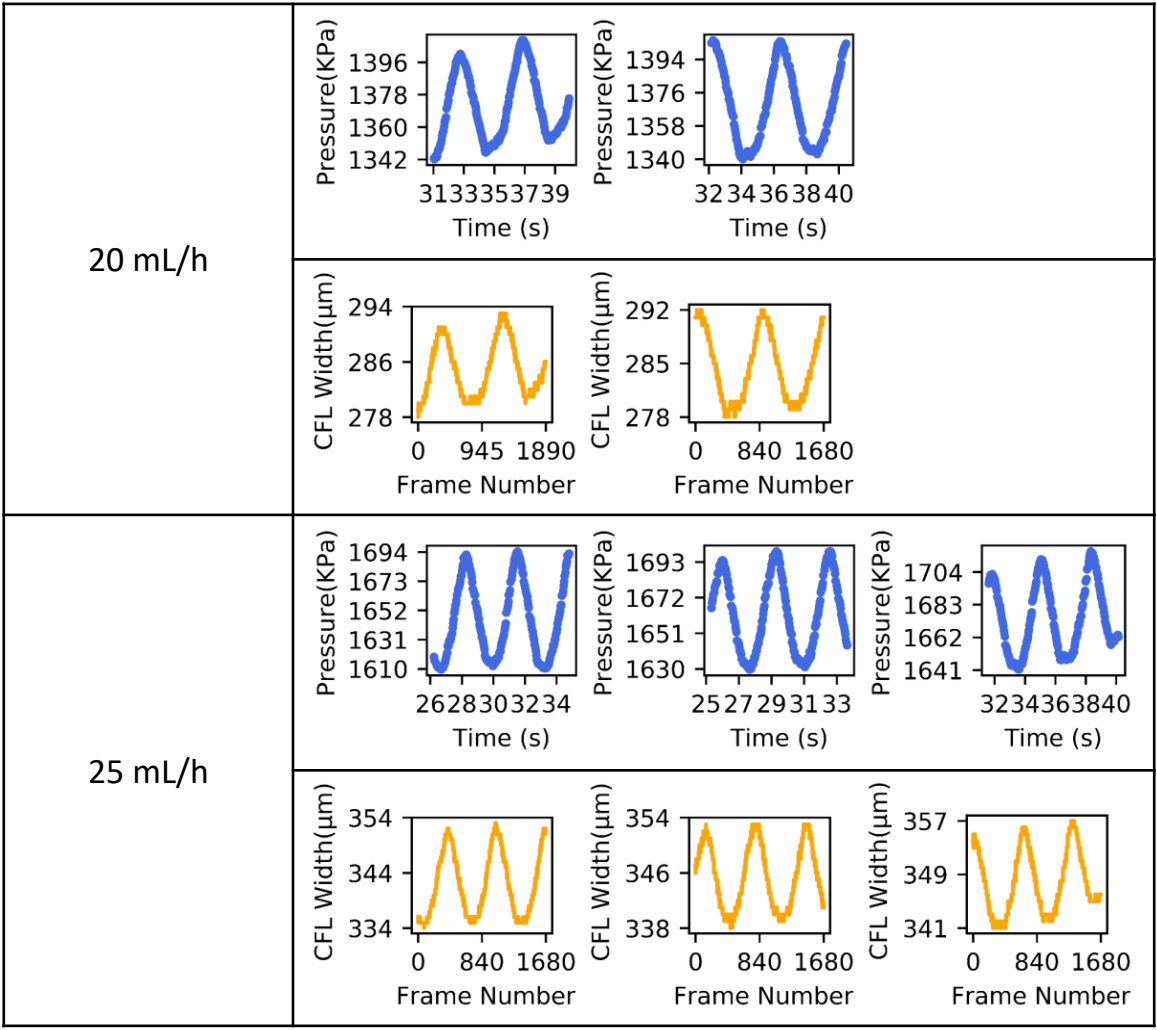 |
| 25 mL/h | |

**Supplementary Table 2:** *Raw data for CFL width (CFZW) (CCRM acquisition) and pressure (LabSmith pressure sensor acquisition) for 20% milk dilution. The dataset were aligned manually, since the image acquisition and the pressure acquisition could not be automatically synchronised. However, the CFL width perfectly mirror the pressure sensor acquisition. The pressure sensor provides 30 sample per second where the CCRM camera is able to capture 210 frames per second.*

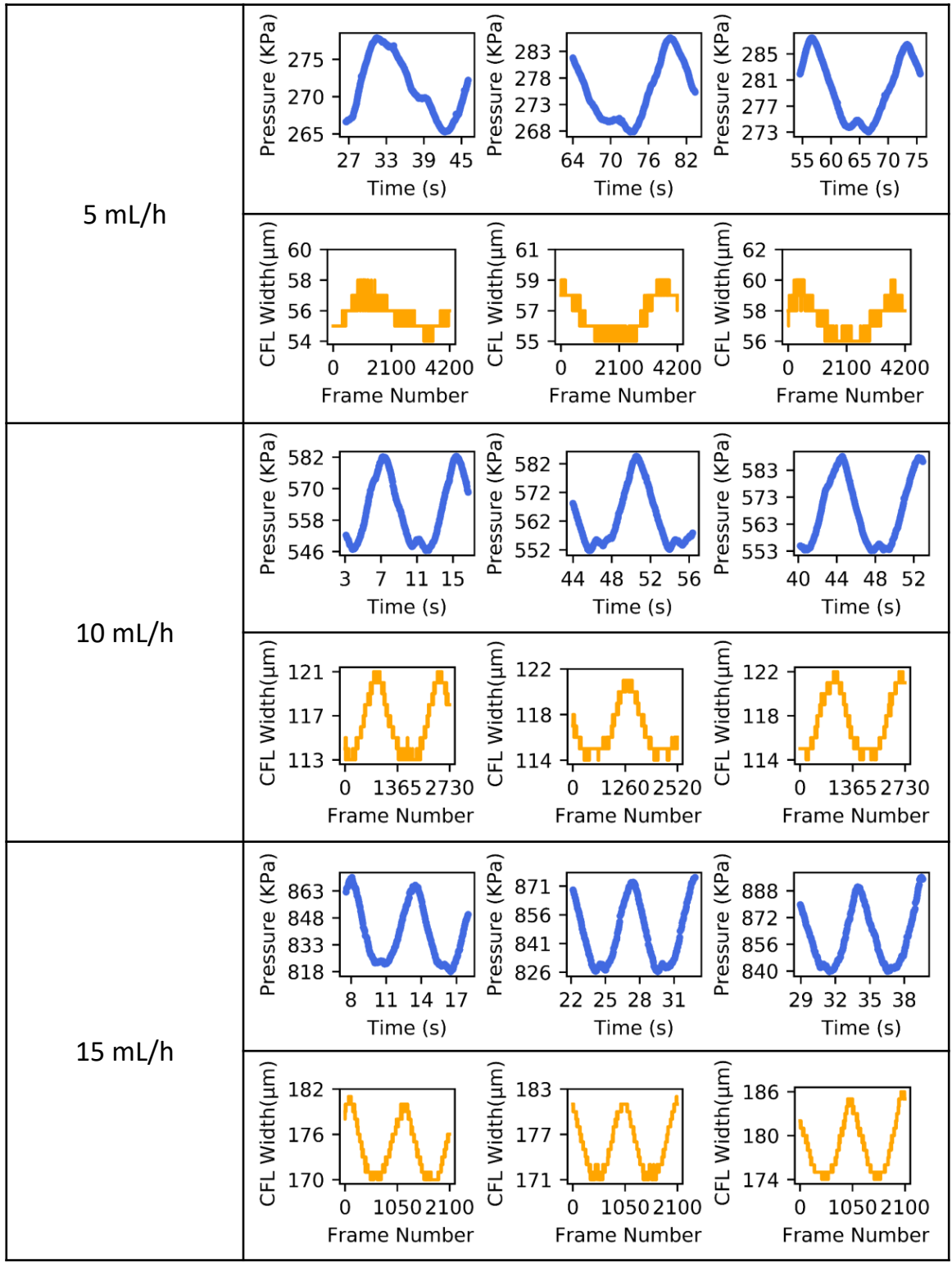

| | |
|---|---|
| 20 mL/h | 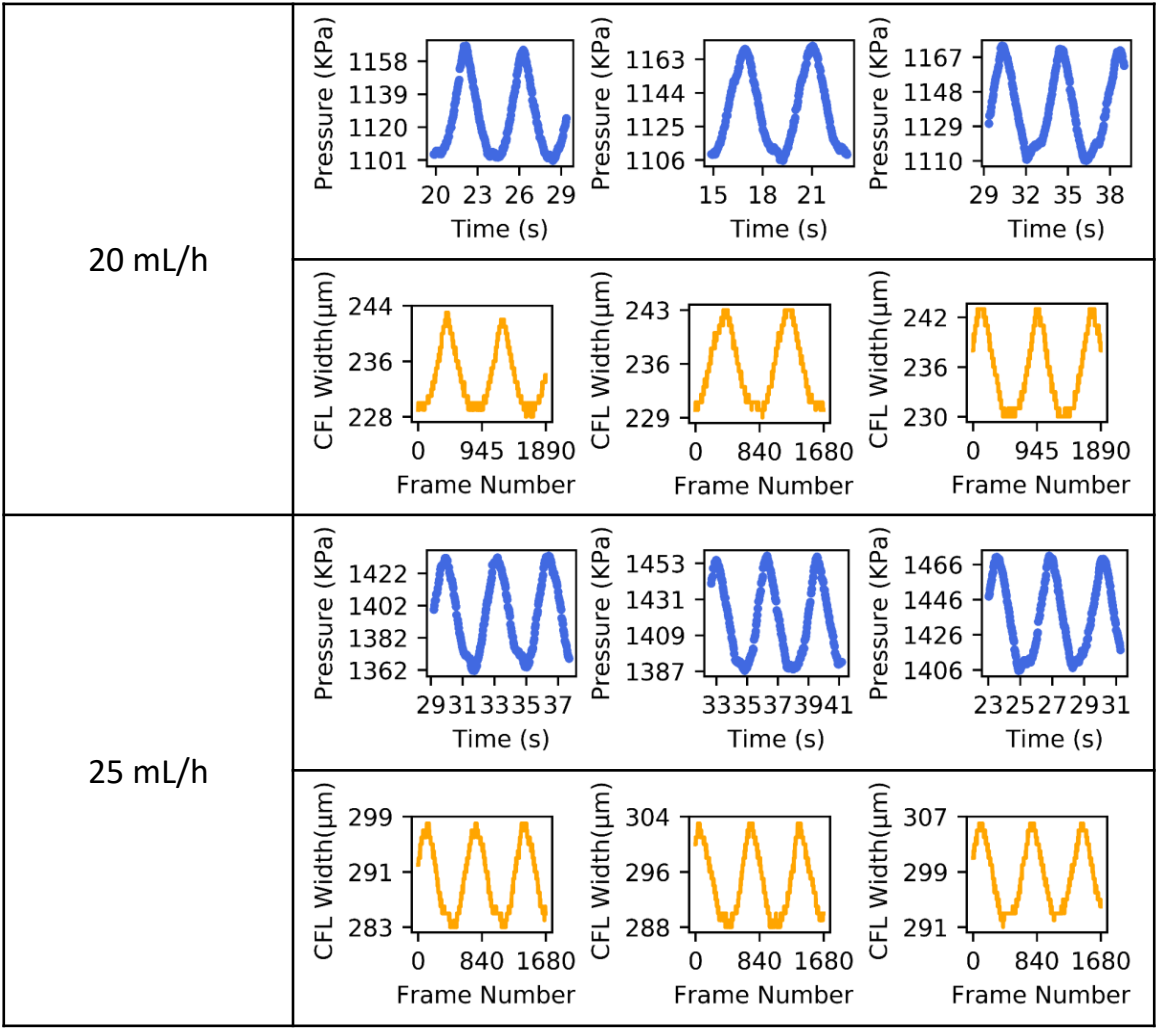 |
| 25 mL/h | |

**Supplementary Table 3:** *Raw data for CFL width (CFZW) (CCRM acquisition) and pressure (LabSmith pressure sensor acquisition) for 10% milk dilution. The dataset were aligned manually, since the image acquisition and the pressure acquisition could not be automatically synchronised. However, the CFL width perfectly mirror the pressure sensor acquisition. The pressure sensor provides 30 sample per second where the CCRM camera is able to capture 210 frames per second.*

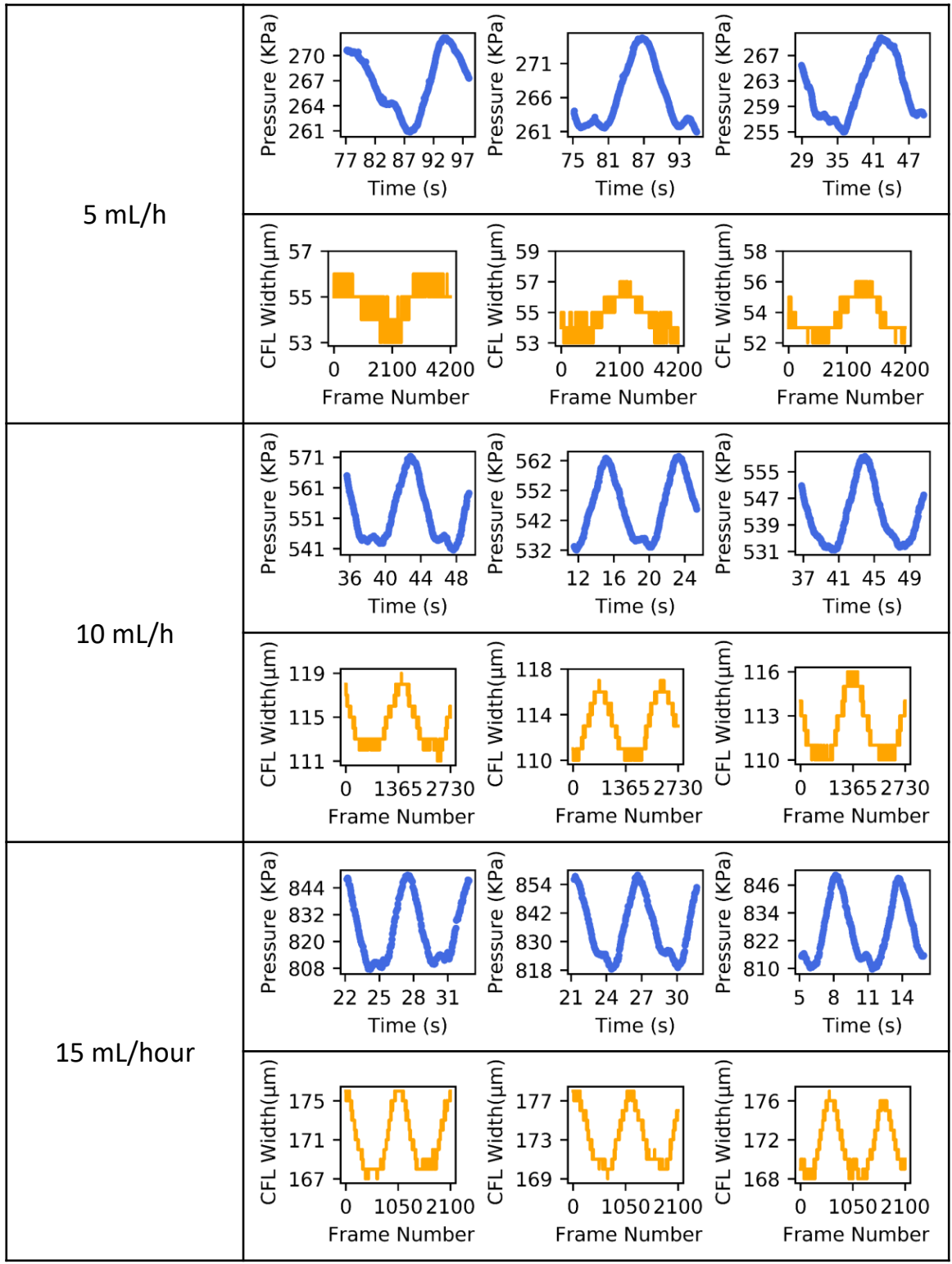

| | |
|---|---|
| 20 mL/h | 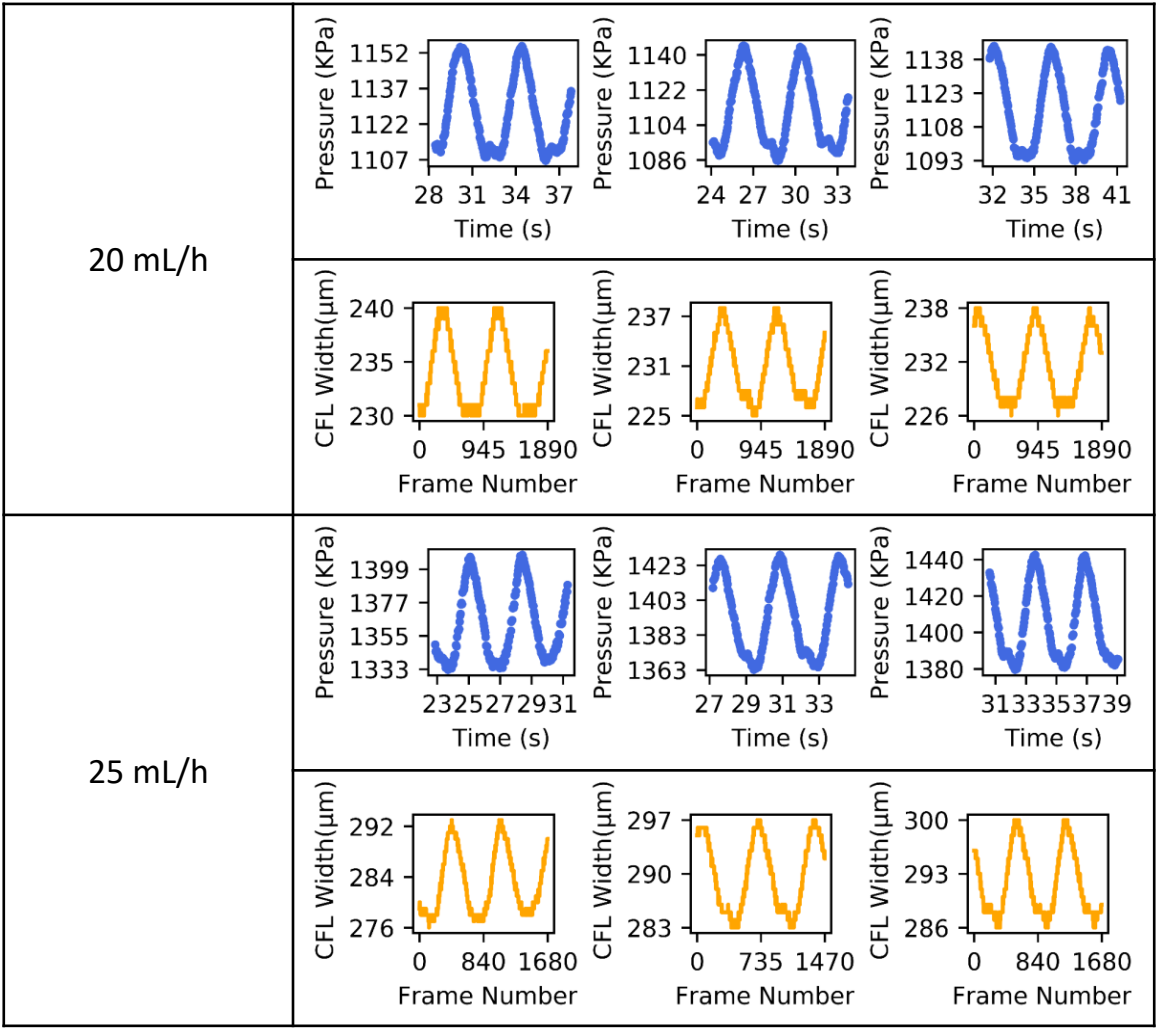 |
| 25 mL/h | |

**Supplementary Table 4:** *Calculated densities, dynamic and kinematic viscosity for the kinematic viscosity experiments*

|  | Density (g/cm³) | Dynamic viscosity (cP) | Kinematic viscosity (cSt) |
|---|---|---|---|
| Water | 0.997 | 1 | 0.997 |
| Milk 50% | 1.015 | 1.0718 | 1.087877 |
| Milk 20% | 1.0042 | 1.1487 | 1.153525 |
| Milk 10% | 1.0006 | 1.414213562 | 1.415062 |
| Glycerol 6.67% | 1.017585669 | 1.2383 | 1.260076 |
| Glycerol 20% | 1.056975506 | 1.985 | 2.098096 |
| Glycerol 33.3% | 1.095588698 | 3.482 | 3.81484 |

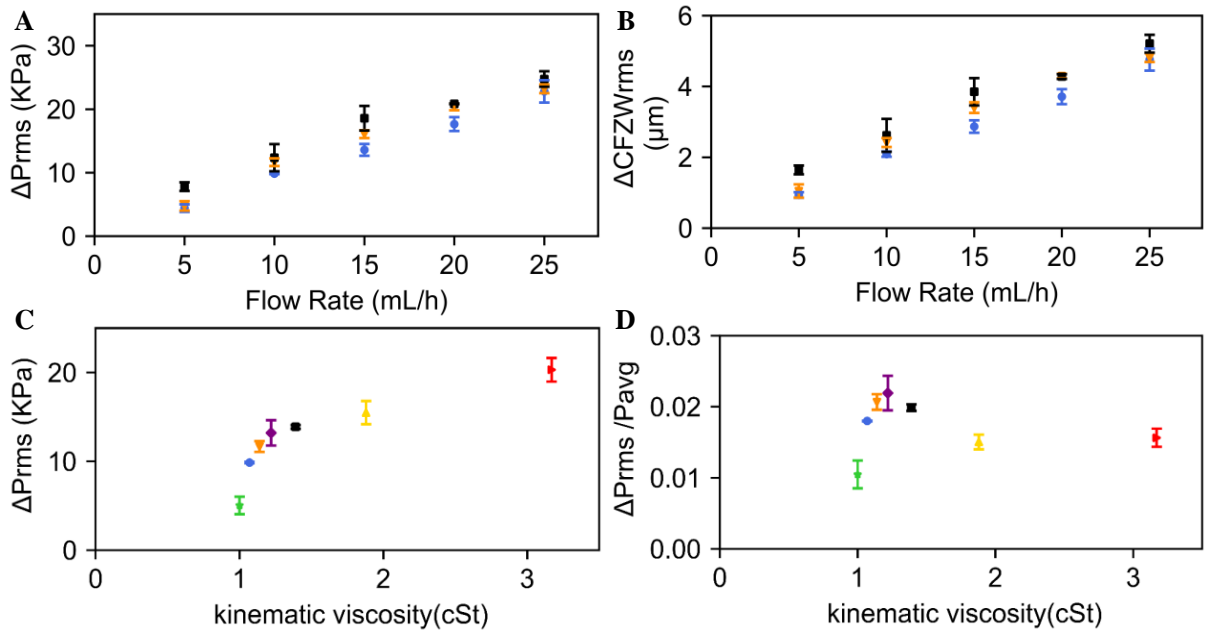

***Supplementary Figure S2:*** *(**A**) Pressure fluctuations at various flow rates and milk dilutions (■50% milk, ▼20% milk, ● 10% milk) (**B**) CFZW fluctuations at various flow rates and milk dilution. (**C**) Pressure fluctuation for samples having different kinematic viscosities at 10 mL/hour (**D**) Relative pressure fluctuations for samples having different kinematic viscosities at 10 mL/hour. Samples: ★water; ● 10% milk ; ▼20% milk; ■50% milk; ◆6.67% glycerol; ▲ 20% glycerol; ▶ 33.33% glycerol.*

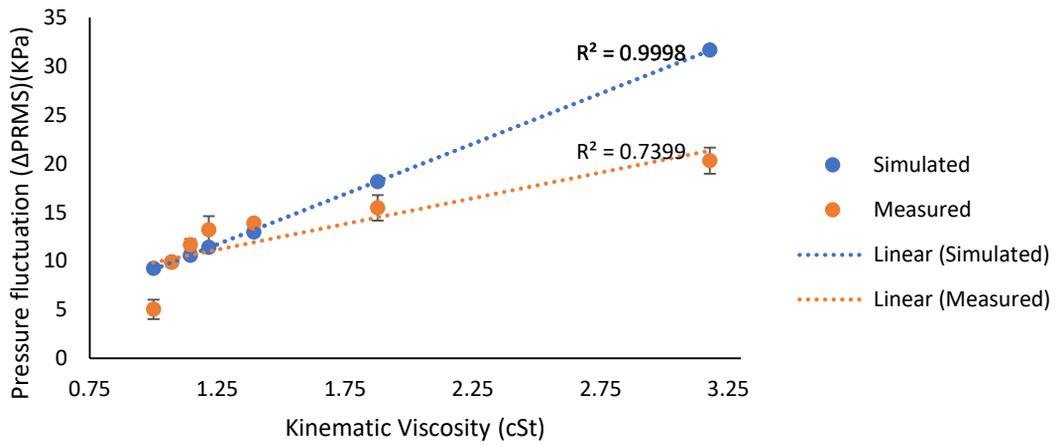
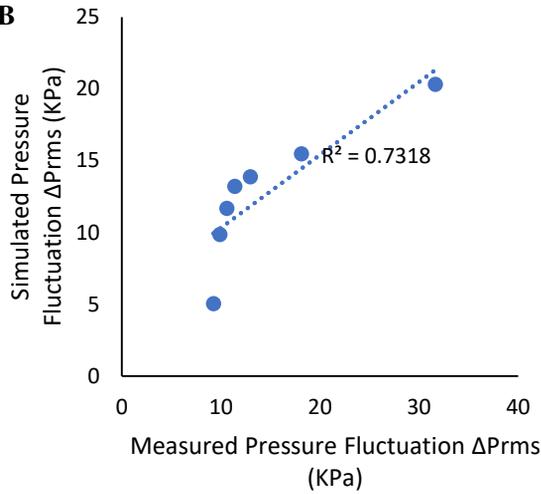
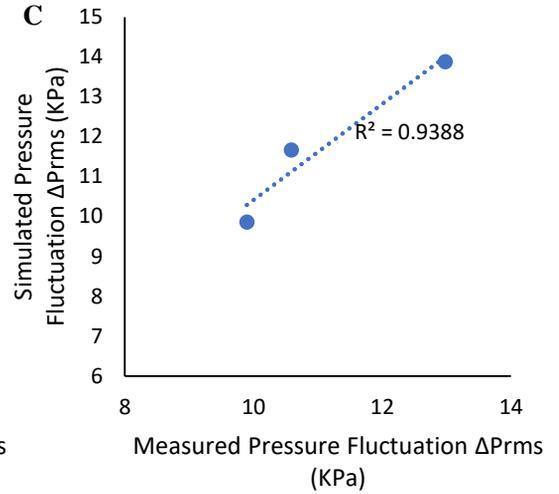

***Supplementary Figure S3:*** *(**A**) Pressure fluctuations vs kinematic viscosity for measured and simulated data. A good agreement exists between the measured and simulated milk samples exists, but deviations for the water sample and glycerol are observed. (**B**) Correlation between simulated and measured fluctuations (all samples) (**C**) Correlation between simulated and measured fluctuations (milk)*

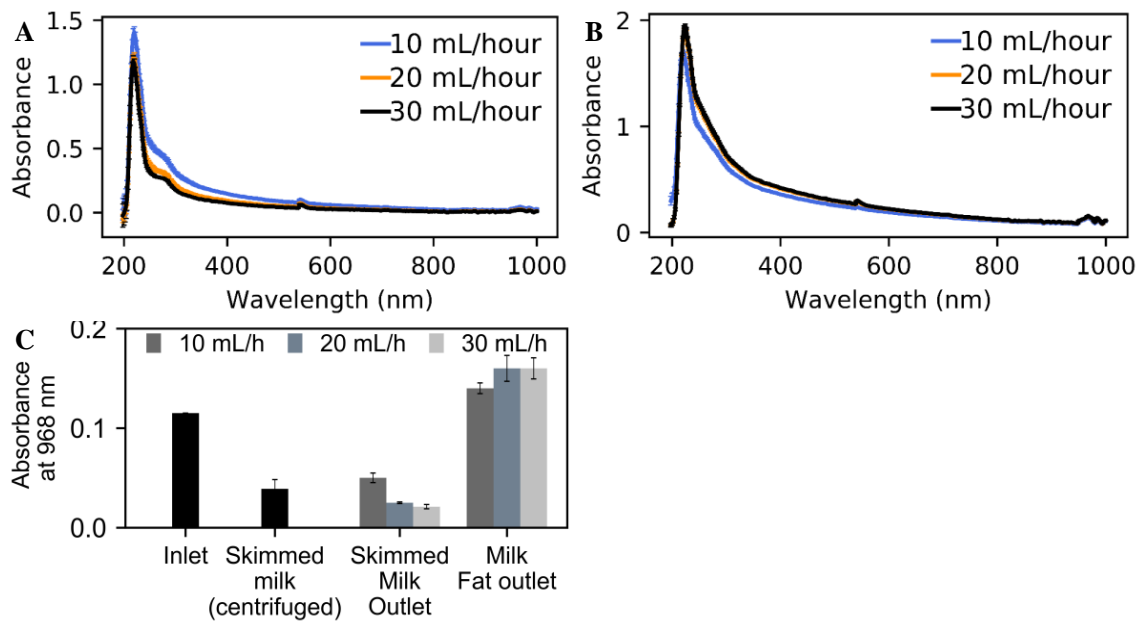

***Supplementary Figure S4:*** *(**A**) Absorbance spectrum of skimmed milk outlet sample after 1:250 dilution with PBS in range (190nm-1000nm) at different flow rates. It can be visible that at 220 nm, sample collected from the low flow rate(10 mL/hour) provides a higher absorbance value compared to the other two (which are almost similar) due to the presence of large number of fat particles. CFL fluctuation is one of the major factors behind this fat particle movement to the skimmed milk outlet. Absorbance peak at 968 nm is very small compared to the 220 nm peak, therefore, is not visible here; (**B**) Absorbance spectrum of Milk fat outlet sample after 1:250 dilution with PBS in range (190nm-1000nm) at different flow rates. At higher flow rates( 20 and 30 mL/hour) most of the fat particles travelled to the milk outlet hence showing higher absorbance peak at 220 nm. 968 nm peak can be visible but difficult to quantify from the figure because of the large difference between the absorbance peak values. (**C**) Absorbance values for the sample collected from Inlet, skimmed milk (Centrifuged 16,000g for 10 minutes), skimmed milk outlet and milk fat outlet at 220 nm*